# High-sensitivity time-resolved intracavity laser Fourier transform spectroscopy with vertical cavity surface emitting multiple quantum well lasers.

Nathalie Picqué, Guy Guelachvili
Laboratoire de Photophysique Moléculaire
Unité Propre du CNRS, Université de Paris-Sud, Bâtiment 350
91405 Orsay-Cedex, France

Alexander A. Kachanov*
Laboratoire de Spectrométrie Physique, Université J. Fourier/C.N.R.S. Grenoble,
BP 87, 38402 Saint Martin d'Hères Cedex, France

Corresponding author :

Nathalie Picqué
Laboratoire de Photophysique Moléculaire
Unité Propre du CNRS, Université de Paris-Sud, Bâtiment 350
91405 Orsay-Cedex, France
Website: http://www.laser-fts.org
email : nathalie.picque@ppm.u-psud.fr

Abstract: Spectra comprised of hundreds of time-components for absorption path lengths up to 130 km have been recorded around 1050 nm by combining two recent techniques, intracavity laser spectroscopy with vertical external cavity surface emitting multiple-quantum-well lasers and time-resolved Fourier transform spectroscopy. A sensitivity of $1 \cdot 10^{-10}$ cm$^{-1}$·Hz$^{-1/2}$ is achieved, for simultaneously acquired $10^4$ spectral elements, three orders of magnitude better than the sensitivity obtained in previous similar experiments. Specific advantages of the method, especially for frequency and intensity metrology of weak absorption transitions, are discussed.

OCIS numbers:
120.0120 Instrumentation, measurement, and metrology
140.5960 Semiconductor lasers
250.7260 Vertical cavity surface emitting lasers
300.6320 Spectroscopy, high-resolution
300.6340 Spectroscopy, infrared

IntraCavity Laser Absorption Spectroscopy (ICLAS) is recognized as one of the most sensitive detection techniques for absorption.[1] Since its introduction in the seventies, the method has taken advantage of various laser gain media and has been applied in many spectroscopic studies, mostly in the visible spectral range. A comprehensive review of the principles and applications of the method is provided in Ref 2. To summarize, the absorbing sample is located inside a laser cavity, whose gain medium is broader than the absorber linewidth. From the beginning of a laser pulse until the observation time $t_g$ (also called generation time), the absorption follows the Lambert-Beer law with path length $L = c\, t_g$, where c is the velocity of light. Sensitivities of the order of $10^{-9}$ cm$^{-1}$ are now routinely reported. Although it is a rather complicated system involving the dynamical evolution of laser modes, a major advantage of intracavity spectroscopy is the simultaneous coverage of a broad spectral domain, similar to the laser gain bandwidth.

The development of ICLAS in the infrared spectral range, despite its obvious benefits, has been hampered by a lack of availability of appropriate lasers (cryogenic operation, high pump power, narrow gain profile, pulsed mode operation, restricted spectral areas) and less efficient spectroscopic approaches. However, the method has been demonstrated with broad gain medium lasers such as $Cr^{4+}$:YAG [3] or Co:$MgF_2$.[4]

Recent progress in semiconductor laser technology has generated new interest in spectroscopic experiments in the infrared. The result: low cost, compact and bright devices are becoming available up to 24 µm.[5] Continuous wave systems with tens of milliwatts power at room temperature exist up to 9 µm,[6] and broadband simultaneous coverage from 6 to 8 µm has been reported.[7] Sensitivities of the order of $10^{-10}$ cm$^{-1}$ have been achieved with ICLAS[8] using Multiple Quantum Wells (MQW) Vertical Cavity Surface Emitting Lasers (VCSEL) around 1 µm. This represents an improvement in sensitivity by four orders of magnitude compared with index guided diode lasers.[9] The technique has been successfully applied in spectroscopic studies.[10] VCSEL can be operated at ~ room temperature up to 3.1 µm.[11] Extending the ICLAS technique to the infrared region should yield many benefits. This extension will be greatly facilitated by the use of Fourier transform spectrometers (FTS) since grating spectrometers with diode arrays are either not available, very expensive, or of poor quality. There are a limited number of ICLAS measurements using FTS that have been performed in the visible region of the spectrum.[12,13,14]

In this letter, we demonstrate that exploiting a recent development made on our time-resolved Fourier transform spectrometer (TRFTS)[15] and combining the two techniques, ICLAS with VCSELs and TRFTS, yields a technique with new and specific advantages. Spectra are obtained around 1 µm for absorption path lengths (APL) of many kilometers using our step-scan interferometer with an intracavity laser absorption set-up based on an optically pumped VCSEL. Below, specific advantages of the method, especially for frequency and intensity metrology of weak absorption transitions, are discussed.

The experimental intracavity set-up, shown in Fig.1, is based on a VCSEL emitting around 1050 nm in an external cavity configuration. The VCSEL structure was grown by molecular beam epitaxy. Its InGaAs quantum wells active medium and its antireflection structure needed for the external cavity configuration are similar to the ones described in Ref 8. The difference is that a metal mirror is placed behind the GaAs/AlAs Distributed Bragg Reflector (DBR) M1 in order to avoid back reflections from the substrate, and the gain structure contains 3 pairs of quantum wells in a 7/2-wavelength cavity. The three-mirror standing-wave cavity is made of (M1), a spherical (150 mm radius curvature) high-reflectivity folding mirror (M2) and a flat output coupler (M3). This L-shaped cavity has a 1-m long arm in which an intracavity cell can be inserted. The semiconductor chip is soldered onto a copper heat sink and is temperature-stabilized by a Peltier element linked to a 30W temperature controller. Tunability is achieved through temperature variation of the chip (– 0.8 nm/°C) and/or by moving the chip and taking advantage of growth rate nonuniformity. Frequency drifts are of the order of 6 MHz/min. [8] The pumping radiation from a continuous fiber-coupled 1-W 830 nm diode laser is focused into the gain region. The pump current is alternately switched above and below the VCSEL threshold (80 mW), therefore initiating the VCSEL spectro-temporal dynamics. Laser pulses have durations chosen between 300 µs and 1 ms. The spectral width of the VCSEL emission is at most 6 nm, near threshold, for low generation times.

Accurate time sampling and full spectral record of the laser pulse are specific advantages brought by TRFTS, with easy a posteriori corrections of the possible frequency and intensity fluctuations. The VCSEL output is sent into the step-scan time-resolved interferometer, equipped with infrasil beam splitter and silicon photodetectors. At a given path difference step, several time samples are taken from a given laser pulse and several pulses are averaged. At the end of the experiment, according to the data acquisition procedure explained in [15], there are as many interferograms as there are time samples . Each time-component interferogram corresponds to a given APL. The time between two consecutive time-components leads to the increment of the APL.

Figure 2 shows an atmospheric absorption spectrum (without the intracavity cell) made of 114 time-components. Typically, from 64 to 512 time-components are recorded. Each time-component is here made of 10,000 samples, obtained after an acquisition time equal to 32 ms. The experiment lasts less than half an hour. The spectral range is limited only by the laser emission. The instrumental spectral resolution is equal to the width of the air-broadened water lines. Thirty-two co-additions have been performed to improve the signal-to-noise ratio (SNR). The electronic bandwidth of the acquisition system is 10 MHz and the time between two consecutive time-components is 3.2 µs. In other words, the APL is increased by ~ 960 m from one spectrum to the next. With a pumping power 10 percents above the laser threshold the APL grows linearly up to 130 km. Due to non-linear mode couplings, this linearity vanishes at higher path lengths. The SNR is better than 100, leading to a sensitivity of $1\ 10^{-10}$ cm$^{-1}$.Hz$^{-1/2}$. Due to the specific interferometric approach,[15] the present results compared with the best previous ICLAS-FTS experiment[13] show an improvement by three orders of magnitude of the sensitivity at 1 second averaging time. Another time-resolved spectrum made of 512 time-components is presented in Figure 3. For the sake of clarity, one component out of 20, from components number 80 to 420 are plotted. Comparison with the HITRAN[16] molecular database reveals the presence of never-measured lines.

In addition to the benefits in the infrared (where other sensitive spectroscopic techniques exist,[17,18] but are not as numerous as in the visible), the ICLAS-TRFTS combination opens new opportunities in the field of wideband sensitive high-resolution spectroscopy. As shown above, a set of time-component spectra with APLs of many kilometers in arithmetic progression is obtained. Due to the particular nature of the step-scan data acquisition, these time-components share many features; the recording time, resolution, SNR, and the entire spectral range with no need of concatenation are identical. The results benefit from the usual advantages of accuracy and resolution of FTS over multichannel grating spectrometers.

We also note, the perfect separation between time and spectral dimensions brings all time-components on exactly the same wavenumber scale. Hence, frequency metrology can be performed with accuracy for profiles over a large range of absorption coefficients. Indeed, absolute frequency standards from saturated absorption laser spectroscopy may be used, since they are generally intense, to calibrate the lower APL spectra. The absolute positions of the weaker lines, well observed in the longer APL time-components, may then be measured with approximately the same accuracy provided by the laser techniques on unsaturated profiles. As far as the spectroscopic parameters are concerned, it is noteworthy that only a very small fraction of the ICLAS papers report intensity determination. Here the sample is observed for all time-components, under identical pressure and temperature conditions, which are easily measured due to the small size of the cell. Moreover, multispectrum-fitting procedures may take advantage of the great number of time-components available. The resulting efficient reduction of systematic errors leads to improved determination of spectroscopic parameters, as recently demonstrated.[19]

This work demonstrates that MQW-VCSEL-based ICLAS experiments coupled with a TRFTS interferometer are well adapted for high-sensitivity laboratory spectroscopy. Moreover, specific benefits of this coupling allow accurate measurements of both frequency and line shape parameters. In addition, this work opens the door to new perspectives of wideband sensitive spectroscopy extended to the infrared region.

The authors are grateful to A. Garnache (CEM2, Montpellier, France) and W.Y. Hwang (AOI, Houston, USA) for the design and growth of the structure, Informed Diagnostics Inc, USA for making the MQW sample available to us. A. Campargue and D. Romanini (LSP Grenoble France) are warmly acknowledged for useful discussions and for lending some material. R.R. Gamache kindly accepted to make English corrections to the manuscript. This research is supported by the "Programme National de Chimie Atmosphérique" (PNCA) du Centre National de la Recherche Scientifique (CNRS). Nathalie Picqué's email address is nathalie.picque@ppm.u-psud.fr
* Present address: PICARRO, Inc., 1050 E. Duane Ave. Suite H, Sunnyvale, CA 94085, USA

Figure Caption

Figure 1: Intracavity set-up. The VCSEL is pumped by a diode laser. See text. Timing procedure for interferometer data acquisition is described in [15].

Figure 2: Time-resolved spectrum of the dynamics of the VCSEL. About 10 absorption lines due to water vapor are observed. Only 2 of them at 1.03852 and 1.03861 µm, connecting the vibrational levels ($v_1, v_2, v_3$) 041 and 000, are in the HITRAN database. Their rotational assignment ($J, K_a, K_c$) is respectively (9,1,9) – (10,1,10) and (9,0,9) – (10,0,10).

Figure 3: Only 1 time-component out of 20 of a 512 time-component spectrum is plotted. Note, the atmospheric water lines become deeper as time grows. The strongest line at 1.04132 µm is assigned to 140 (5,0,5) <– 000 (6,1,6).

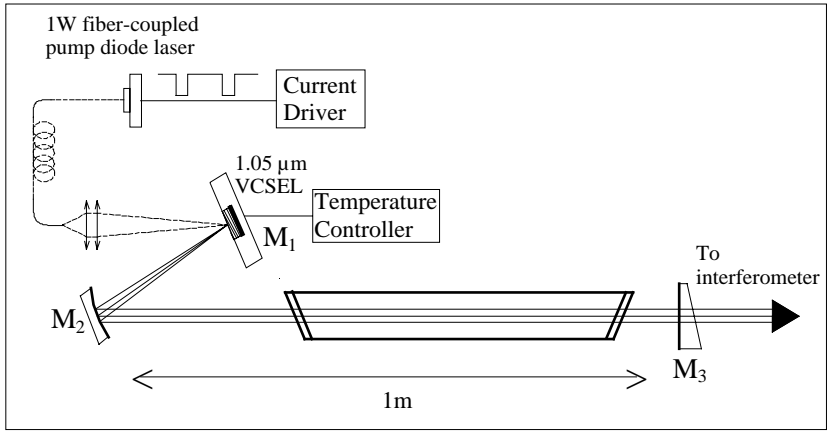

**Figure 1**

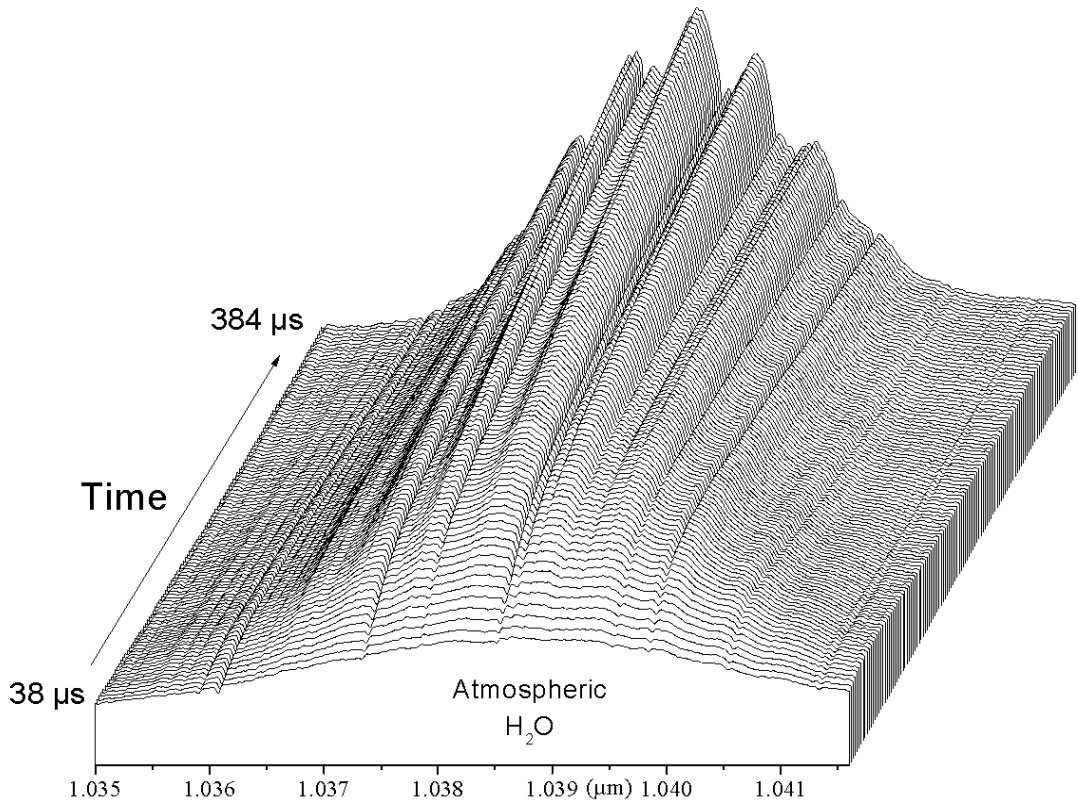

Figure 2

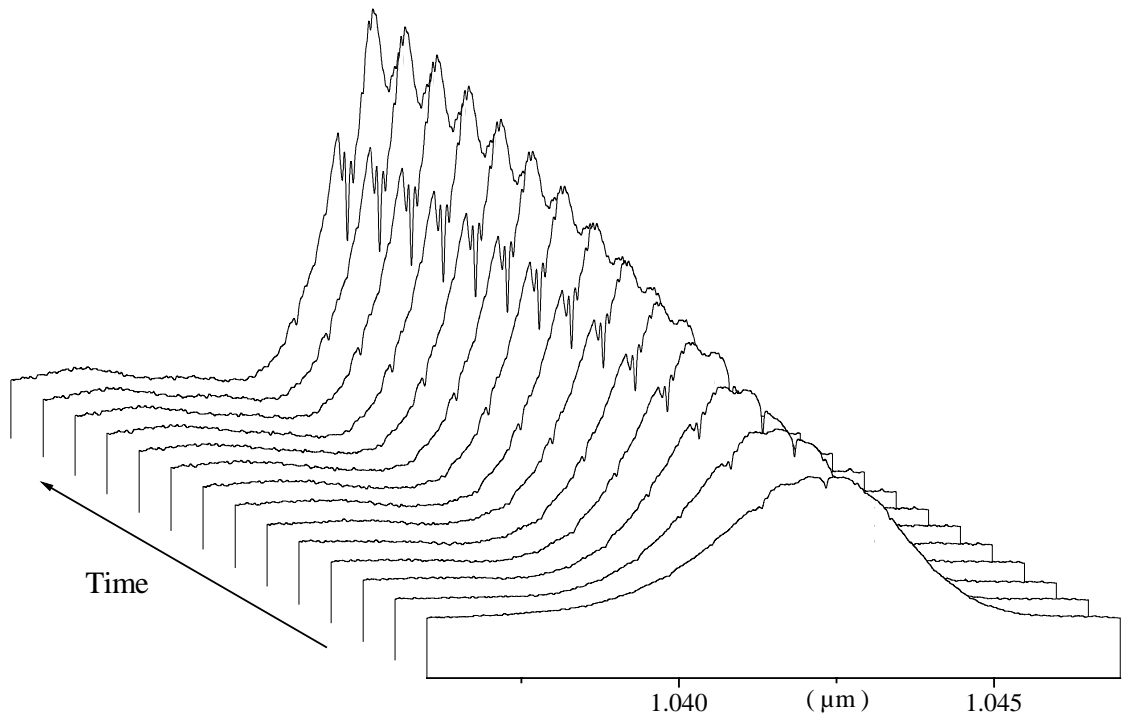

Figure 3